\documentclass[aps,amssymb,prl,twocolumn]{revtex4}
\usepackage{graphicx}
\usepackage{epsfig,amsmath}

\begin{document}

%\title{Transition effects induced by frequency-dependent third cumulant of current fluctuations on a small quantum system}
\title{Mesoscopic photon heat transistor}

\author{Teemu Ojanen$^1$}
\email[Correspondence to ]{teemuo@boojum.hut.fi}
\author{Antti-Pekka Jauho$^{2,3}$}
\affiliation{$^1$ Low Temperature Laboratory, Helsinki University of
Technology, P.~O.~Box 2200, FIN-02015 HUT, Finland }
\affiliation{$^2$ MIC - Department of Micro and Nanotechnology,
NanoDTU, Technical University of Denmark, Orsted Plads, Bldg, 345E,
DK-2800 Kgs, Lyngby, Denmark } \affiliation{$^3$ Laboratory of
Physics, Helsinki University of Technology, P.~O.~Box 1100,
FIN-02015 HUT, Finland }

\date{\today}
\begin{abstract}
We show that the heat transport between two bodies, mediated by
electromagnetic fluctuations, can be controlled with an intermediate
quantum circuit - leading to the device concept Mesoscopic Photon
Heat Transistor (MPHT).  Our theoretical analysis is based on a
novel Meir-Wingreen-Landauer type of conductance formula, which
gives the photonic heat current through an arbitrary circuit element
coupled to two dissipative reservoirs at finite temperatures. As an
illustration we present an exact solution for the case when the
intermediate circuit can be described as an electromagnetic
resonator.  We discuss in detail how the MPHT can be implemented
experimentally in terms of a flux-controlled SQUID circuit.
%We study theoretically heat transport mediated by electromagnetic
%fluctuations in nanostructures. Applying nonequilibrium Green's
%function methods we derive Meir-Wingreen-Landauer formula for the
%photonic heat current through an arbitrary circuit element coupled
%to two linear dissipative reservoirs at finite temperatures. The
%transport problem is solved exactly in the case where the mediating
%circuit is an electromagnetic resonator. By externally modulating
%the resonator properties one can control the heat current through
%the structure. We discuss in detail how the phenomenon can be
%experimentally realized in a flux-controlled SQUID circuit.
\end{abstract}
\pacs{PACS numbers: } \bigskip

\maketitle

Problems involving interactions between quantum particles and
electromagnetic fields have a long and rich history. During the last
decades developments in mesoscopic physics have provided
unprecedented possibilities to engineer the electron-photon
interactions. A major advantage in mesoscopic systems is their
versatility which allows a high-level of control in their design and
operation. Recent advances in this field include impressive quantum
state manipulations involving single photons in circuit cavity QED
experiments \cite{houck} and a demonstration of single-channel
photon heat transport \cite{meschke}.

At low temperatures, when phonon modes become effectively frozen,
photonic heat conduction becomes the dominant channel for thermal
transport \cite{schmidt,meschke}. In this Letter we study photonic
heat transport in a structure consisting of two reservoirs coupled
via an intermediate electric circuit. The reservoirs are assumed to
behave as linear dissipative circuit elements and are thereby fully
characterized by their response functions and temperatures.
Furthermore, at low temperatures the wavelengths of relevant field
fluctuations are much longer than a typical system size so the
reservoirs can be effectively considered as lumped elements. Within
these assumptions we apply the Caldeira-Leggett procedure and model
the reservoirs as continuous distributions of harmonic oscillators.
By applying nonequilibrium Green's function methods we derive a
formally exact Meir-Wingreen-Landauer- type formula \cite{meir} for
the heat current through the structure. Our formula involves the
{\it noise power} of the intermediate circuit, in the presence of
the coupling to the leads, and serves as a general starting point
for solving the heat transport problem. We find an exact solution
for the heat current flowing through an electromagnetic resonator
circuit and show explicitly that, in analogy with the transistor
effect in charge transport problems, the heat flow through the
structure can be modulated by applying an external control to the
middle circuit. We suggest that an experimental demonstration of the
heat-transistor action can be achieved by using a Superconducting
QUantum Interference Device (SQUID) circuit as the tunable
resonator. Thus, the magnetic flux controlled heat current is a
photonic analogue to the gate voltage controlled electronic heat
current recently demonstrated in Ref.~\cite{saira}.

\begin{figure}[h]
\centering
\includegraphics[width=0.9\columnwidth]{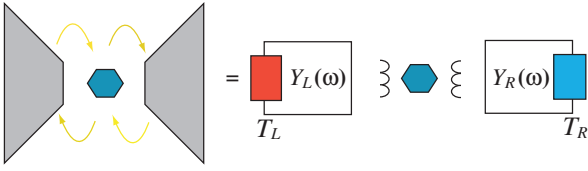}
\caption{Temperature gradient between the left and right reservoirs
induces photonic heat current through an arbitrary quantum circuit
coupled to them. Heat flows from left to right when $T_L>T_R$. The
reservoirs are assumed to behave as linear dissipative elements and
they couple only through the middle circuit. In principle a direct
coupling between the reservoirs always exists but in practice it can
be made negligible by an appropriate sample design.} \label{setuppi}
\end{figure}

Now we turn to the technical derivation of the general formula for
the heat current in the system consisting of a left and a right
reservoir, and an arbitrary quantum circuit between them
(Fig.~\ref{setuppi}). We treat the problem by employing a
nonequilibrium Green's function method analogous to those used
earlier in electron and heat transport problems
\cite{meir,haug,wang,yamamoto,ojanen}. The reservoirs are described
by quadratic boson fields and, according to the Caldeira-Leggett
prescription, can be thought of as arbitrary linear electric
circuits by choosing specific distributions of frequencies
$\omega_j$ and couplings $g_j$ (introduced below)
\cite{devoret,legget}. The total Hamiltonian is assumed to be of the
form $H=H_L+H_R+H_{M}+H_{C}$, where
\begin{align}
H_{L/R}=\sum_{j\in
L/R}\hbar\omega_j(\hat{a}_j^{\dagger}\hat{a}_j+1/2),
\end{align}
and the inductive coupling term is
\begin{align}
H_{C}=M\hat{I}\left(\hat{i}_{L}+ \hat{i}_{R}\right),
\end{align}
which involves the current operators for the central region
$\hat{I}$ and for the reservoirs $\hat{i}_{L/R}=\sum_{j\in
L/R}g_j(\hat{a}_j+\hat{a}_j^{\dagger})$, respectively. The specific
form of the Hamiltonian and the current operator of the middle
region do not need to be specified at this point (below we shall
treat a specific example).
%The reservoir operators obey the usual
%Boson commutation relations $[\hat{a}_i,
%\hat{a}_j^{\dagger}]=\delta_{ij}$.
The mutual inductances between the middle circuit and the leads are
assumed to be equal, though this is not necessary in the following
derivation. A capacitive coupling between the reservoirs and the
middle system can be treated in close analogy with the inductive
coupling studied here \cite{note}.
% \footnote{A direct coupling
%between the reservoirs could easily be included in the formalism,
%but since it is much smaller due to the device geometry, we will not
%discuss it here.}

The reservoir Hamiltonians $H_{L/R}$ do not commute with the total
Hamiltonian $H$, thus giving a rise to an energy flow in the
structure. This energy flow is characterized by a heat current
$J_{L/R}$ defined as
\begin{align}\label{cur}
J_{L/R}(t)&=\langle\dot{H}_{L/R}\rangle=iM\sum_{j\in
L/R}\left[g_j\omega_j\langle \hat{a}_j(t)\hat{I}(t)\rangle-\mathrm{h.c.}\right]\nonumber\\
&=-2M\mathrm{Re}\sum_{j\in L/R}g_j\omega_jG_j^{<}(t,t),
\end{align}
where $G_j^{<}(t,t')\equiv -i\langle
\hat{a}_j(t')\hat{I}(t)\rangle$.
%\begin{align}\label{green}
%G_j^{<}(t,t')\equiv -i\langle \hat{a}_j(t')\hat{I}(t)\rangle.
%\end{align}
The transport problem is reduced to finding the lesser Green's
function $G_j^{<}(t,t')$ which can be derived by the
equation-of-motion technique \cite{haug, bruus}. Following the
standard prescription \cite{haug}, we first consider the equilibrium
zero-temperature time-ordered correlation functions, which obey
\begin{equation}\label{time1}
(i\partial_{t'}-\omega_j)\langle T[\hat{a}_j(t')\hat{I}(t)]
\rangle=\frac{Mg_j}{\hbar}\langle T[\hat{I}(t')\hat{I}(t)] \rangle,
\end{equation}
or, after a formal integration,
\begin{equation}\label{time2}
\langle T[\hat{a}_j(t')\hat{I}(t)]\rangle=\frac{Mg_j}{\hbar}\,\int
dt_1\langle T[\hat{I}(t)\hat{I}(t_1)]\rangle D_j(t_1-t'),
\end{equation}
where $D_j(t_1-t')$ is the free reservoir Green's function.
% at $T=0$.
In nonequilibrium, this equation holds on the Keldysh contour, and
using the analytical continuation rules known as Langreth's theorem
\cite{haug}, we obtain
\begin{align}\label{val}
G_j^<(t,t')=&\frac{Mg_j}{\hbar}\,\int
dt_1\left[\langle\hat{I}(t)\hat{I}(t_1)\rangle^{r}D_j^<(t_1-t')+\right.\nonumber\\
&+\left.\langle\hat{I}(t)\hat{I}(t_1)\rangle^{<}D_j^a(t_1-t')\right],
\end{align}
where the superscripts $r$, $a$ and $<$ stand for "retarded",
"advanced" and "lesser", respectively.
%and $D$ stands for free boson Green's
%functions.
Explicitly, the current correlation functions are
$\langle\hat{I}(t)\hat{I}(t')\rangle^{r}=-i\theta(t-t')\langle[\hat{I}(t),\hat{I}(t')]\rangle$
and $\langle\hat{I}(t)\hat{I}(t')\rangle^{<}=
-i\langle\hat{I}(t')\hat{I}(t)\rangle$. In a steady state
$G_j^<(t,t')=G_j^<(t-t')$, and it is convenient to introduce the
Fourier transform:
\begin{equation}\label{fou}
\!\!\!G_j^<(\omega)=\frac{Mg_j}{\hbar}\left[\langle\hat{I}\hat{I}\rangle^{r}(\omega)D_j^<(\omega)+
\langle\hat{I}\hat{I}\rangle^{<}(\omega)D_j^a(\omega)\right],
\end{equation}
where $D_j^a(\omega)=1/(\omega-\omega_j-i\eta)$,
$D_j^<(\omega)=-i2\pi n(\omega_j)\delta(\omega-\omega_j)$, and
$n(\omega)$ is the Bose function. We thus obtain
\begin{align}\label{current}
J_L=2\sum_{j\in
L}\frac{M^2g_j^2\omega_j}{2\pi\hbar}&\int_{-\infty}^{\infty}\!\!\!\!d\omega
\left[ -\mathrm{Im}\langle\hat{I}\hat{I}\rangle^{r}(\omega)2\pi
n(\omega_j) \delta(\omega-\omega_j)\right.\nonumber\\
&\left.+\mathrm{Im}\langle\hat{I}\hat{I}\rangle^{<}(\omega)\pi
\delta(\omega-\omega_j)\right].
\end{align}
To make a connection to quantities with a clear physical
interpretation, it is useful to express Eq.~(\ref{current}) in terms
of the noise power. The noise power for an observable $\hat{A}$ is
defined as
$S_A(\omega)=\int_{-\infty}^{\infty}dt\,e^{i\omega(t-t')}\langle\hat{A}(t)\hat{A}(t')\rangle$,
%\begin{equation}\label{noise1}
%S_I(\omega)=\int_{-\infty}^{\infty}dt\,e^{i\omega(t-t')}\langle\hat{I}(t)\hat{I}(t')\rangle,
%\qquad
%S_{i_{L/R}}(\omega)=\int_{-\infty}^{\infty}dt\,e^{i\omega(t-t')}\langle\hat{i}_{L/R}(t)\hat{i}_{L/R}(t')\rangle.
%\end{equation}
so the current noise in the left lead is
\begin{align}\label{noise2}
S_{i_L}(\omega)%=\sum_{j\in L}S_{i_j}(\omega)\equiv
%\int_{-\infty}^{\infty}dt\,e^{i\omega(t-t')}\sum_{j\in
%L}\langle\hat{i}_j(t)\hat{i}_j(t')\rangle \nonumber\\ &=
%\int_{-\infty}^{\infty}dt\,e^{i\omega(t-t')}\sum_{j\in L} \langle
%T_j(a_j(t)+a_j^{\dagger}(t'))T_j(a_j(t)+a_j^{\dagger}(t'))\rangle
=\sum_{j\in
L}&g_j^2\left[n(\omega_j)2\pi\delta(\omega+\omega_j)+\right.\nonumber\\
&\left.+(n(\omega_j)+1)2\pi\delta(\omega-\omega_j) \right].
\end{align}
%so we have relations
%\begin{align}\label{noise3}
%& T_j^2\omega_j2\pi\delta(\omega-\omega_j)n(\omega_j)=\omega S_{i_j}(-\omega)\theta(\omega)\nonumber\\
%&T_j^2\omega_j2\pi\delta(\omega-\omega_j)=\omega\left[S_{i_j}(\omega)-S_{i_j}(-\omega)\right]\theta(\omega).
%\end{align}
In terms of the noise power Eq.~(\ref{current}) becomes
\begin{align}\label{current2}
%&J_L=\sum_{j\in
%L}\frac{2M^2}{2\pi\hbar}\int_{-\infty}^{\infty}d\omega \left[
%-\mathrm{Im}\langle\hat{I}\hat{I}\rangle^{r}(\omega)\omega
%S_{i_j}(-\omega)\theta(\omega)
%+\mathrm{Im}\langle\hat{I}\hat{I}\rangle^{<}(\omega)\frac{\omega}{2}\left[S_{i_j}(\omega)-S_{i_j}(-\omega)\right]\theta(\omega)
%\right]=\nonumber \\
J_L=2M^2&\int_{0}^{\infty}\frac{d\omega\omega}{2\pi\hbar} \left[
-\mathrm{Im}\langle\hat{I}\hat{I}\rangle^{r}(\omega)
S_{i_L}(-\omega)\right.\nonumber\\
&\left.+\mathrm{Im}\langle\hat{I}\hat{I}\rangle^{<}(\omega)\frac{1}{2}\left(S_{i_L}(\omega)-S_{i_L}(-\omega)\right)
\right].
\end{align}
Also the $\hat{I}$-correlation functions can be written in terms of
noise power:
$-\mathrm{Im}\langle\hat{I}\hat{I}\rangle^{r}(\omega)=\frac{1}{2}(S_I(\omega)-S_I(-\omega))$
and
$\mathrm{Im}\langle\hat{I}\hat{I}\rangle^{<}(\omega)=-S_I(-\omega)$,
which yields
\begin{equation}\label{current3}
J_L%=M^2\int_{0}^{\infty}\frac{d\omega\omega}{2\pi\hbar} \left[
%\left(S_I(\omega)-S_I(-\omega)\right)S_{i}^L(-\omega)
%-S_I(-\omega)\left(S_{i}^L(\omega)-S_{i}^L(-\omega)\right)
%\right]\nonumber \\
=M^2\int_{0}^{\infty}\frac{d\omega\omega}{2\pi\hbar}
\left[S_I(\omega)S_{i_L}(-\omega) -S_I(-\omega)S_{i_L}(\omega)
\right].%=M^2\int_{-\infty}^{\infty}\frac{d\omega\omega}{2\pi\hbar}S_I(\omega)S_{i_L}(-\omega).
\end{equation}
A similar expression holds for $J_R$;  in steady-state situations
$J\equiv J_L=-J_R$. Formulas (\ref{current2}) and (\ref{current3})
contain the free reservoir noise functions which can be obtained
straightforwardly from the admittances $Y_{L/R}(\omega)$ and
temperatures $T_{L/R}$ by applying the Fluctuation-Dissipation
theorem \cite{devoret}:
\begin{equation}\label{FD}
S_{i_{L/R}}(\omega)=\mathrm{Re}[Y_{L/R}(\omega)]\hbar\omega\left[\mathrm{coth}(\beta_{L/R}\hbar\omega/2)+1\right].
\end{equation}
With the help of Eq.~(\ref{FD}), the heat current (\ref{current3})
can be written as
\begin{align}\label{current4}
J_L=&\int_{0}^{\infty}\frac{d\omega\omega^2M^2}{2\pi} \left\{
2\left[S_I(\omega)-S_I(-\omega)\right]\mathrm{Re}[Y_{L}(\omega)]n_L(\omega)\right.\nonumber\\
&\left.-S_I(-\omega)2\mathrm{Re}[Y_{L}(\omega)]\right\}
\end{align}
If the lead admittances share the same frequency dependence,
$Y_L(\omega)=cY_R(\omega)$ with some constant $c$, the stationary
heat current can be expressed as $J=J_L/(c+1)-cJ_R/(c+1)$ and cast
into the Landauer form:
\begin{align}\label{landauer}
J=&M^2\int_{0}^{\infty}\frac{d\omega\omega^2}{2\pi} [
S_I(\omega)-S_I(-\omega)] \nonumber\\
&\times\frac{2\mathrm{Re}[Y_{L}(\omega)]\mathrm{Re}[Y_{R}(\omega)]}{\mathrm{Re}[Y_{L}(\omega)]+\mathrm{Re}[Y_{R}(\omega)]}
[n_L(\omega)-n_R(\omega)].
\end{align}
%where $n_{L/R}(\omega)$ are Bose functions at reservoir temperatures
%\cite{meir}.
Equation (\ref{landauer}) is the main formal result of
this Letter. We emphasize the strong analogue between this result
and the Meir-Wingreen conductance formula: here the ``bosonic
thermal window" $(n_L-n_R)$ replaces the ``fermionic voltage window"
$(n_L^F-n_R^F)$, the real part of the admittance plays the role of
the line-width function $\Gamma_{L/R}$, and the noise power
$S_I(\omega)-S_I(-\omega)$ of the central region, which needs to be
evaluated separately, replaces the central region spectral function.
%The simple forms of Eqs.~(\ref{current4}) and
%(\ref{landauer}) are deceptive; the noise power of the intermediate
%system $S_I(\omega)$ needs to be calculated separately.
Since  $S_I$ contains information of the internal dynamics of the
middle region {\it in the presence of coupling to the reservoirs},
its evaluation may be difficult indeed, and only in special cases
analytic progress can be expected.
%In the weak coupling limit
%the effects of the leads on $S_I(\omega)$ may be ignored, but
%generally finding $S_I(\omega)$ is a major task and can be carried
%out exactly only in special cases.
In the following we calculate $S_I(\omega)$ for an electromagnetic
resonator and illustrate how the reservoir admittances come into
play through the self-energy of the middle circuit.

\begin{figure}[h]
\centering
\includegraphics[width=0.9\columnwidth]{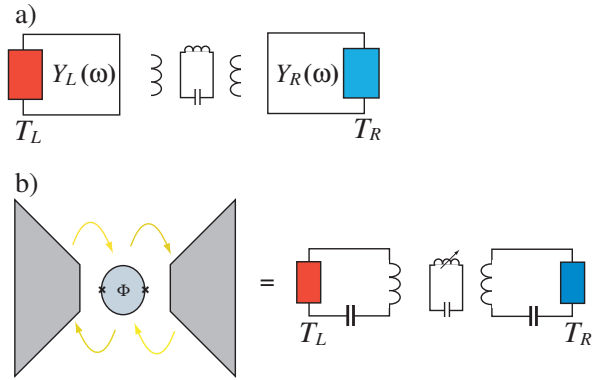}
\caption{An electromagnetic resonator as the intermediate circuit
(a)). In b) the mediate resonator circuit with a tunable inductance
is realized by a dc-SQUID with an external magnetic flux bias
$\Phi$. The external flux can be used to modulate the heat current
through the structure and plays an analogous role to the gate
voltage in electronic transistor.} \label{res}
\end{figure}

Suppose now that the mediating quantum circuit is an electromagnetic
resonator with inductance $L$ and capacitance $C$ (see
Fig.~\ref{res} a). The central region Hamiltonian then takes the
form $H_M=\hbar\omega_0(\hat{b}^{\dagger}\hat{b}+\frac{1}{2})$, and
the current operator is $\hat{I}=I_0(\hat{b}+\hat{b}^{\dagger})$,
where $I_0=\sqrt{\hbar\omega_0/2L}$ and $\omega_0=1/\sqrt{LC}$;
$\hat{b}$, $\hat{b}^{\dagger}$ are bosonic creation and annihilation
operators, $[\hat{b},\hat{b}^{\dagger}]=1$. To evaluate
$(\ref{landauer})$ one needs to find the retarded function
$\langle\hat{I}(t)\hat{I}(t')\rangle^{r}=-i\theta(t-t')\langle[\hat{I}(t),\hat{I}(t')]\rangle$
which can be expressed as sum of four different retarded functions
\begin{align}\label{ret1}
\langle\hat{I}\hat{I}\rangle^{r}(\omega)=I_0^2\left[ \langle
\hat{b}\hat{b}^{\dagger}\rangle^r(\omega)+\langle
\hat{b}^{\dagger}\hat{b}\rangle^r(\omega) +\right. \nonumber\\
\left.\langle \hat{b}\hat{b}\rangle^r(\omega)+ \langle
\hat{b}^{\dagger}\hat{b}^{\dagger}\rangle^r(\omega) \right].
\end{align}
%where the functions on the right-hand side are Fourier transforms of
%\begin{align}\label{ret}
%\langle \hat{b}(t)\hat{b}^{\dagger}(t')\rangle^r
%&\equiv-i\theta(t-t')\langle[\hat{b}(t),\hat{b}^{\dagger}(t')]\rangle,\nonumber\\
%\langle \hat{b}^{\dagger}(t)\hat{b}(t')\rangle^r
%&\equiv-i\theta(t-t')\langle[\hat{b}^{\dagger}(t),\hat{b}(t')]\rangle,\nonumber\\
%\langle \hat{b}(t)\hat{b}(t')\rangle^r
%&\equiv-i\theta(t-t')\langle[\hat{b}(t),\hat{b}(t')]\rangle,\nonumber\\
%\langle \hat{b}^{\dagger}(t)\hat{b}^{\dagger}(t')\rangle^r
%&\equiv-i\theta(t-t')\langle[\hat{b}^{\dagger}(t),\hat{b}^{\dagger}(t')]\rangle.
%\end{align}
Note the presence of the anomalous functions $\langle
\hat{b}(t)\hat{b}(t')\rangle^r$ and $\langle
\hat{b}^{\dagger}(t)\hat{b}^{\dagger}(t')\rangle^r$; they are
important and come into play because the interaction term also
includes the non-rotating-wave terms $\hat{a}_j\hat{b}$ and
$\hat{a}_j^{\dagger}\hat{b}^{\dagger}$. By using an
equation-of-motion technique \cite{haug,bruus} we find a closed set
of equations
\begin{align}\label{eqof3}
(\omega-\omega_0+i\eta)\langle \hat{b}\hat{b}^{\dagger}\rangle^r & =
1+ \Sigma^r(\omega) \left[\langle \hat{b}\hat{b}^{\dagger}\rangle^r+
\langle
\hat{b}^{\dagger}\hat{b}^{\dagger}\rangle^r\right]\nonumber\\
(\omega+\omega_0+i\eta)\langle
\hat{b}^{\dagger}\hat{b}^{\dagger}\rangle^r & = -\Sigma^r(\omega)
\left[\langle \hat{b}\hat{b}^{\dagger}\rangle^r+ \langle
\hat{b}^{\dagger}\hat{b}^{\dagger}\rangle^r\right]\nonumber\\
(\omega+\omega_0+i\eta)\langle \hat{b}^{\dagger}\hat{b}\rangle^r & =
-1- \Sigma^r(\omega)\left[\langle \hat{b}\hat{b}\rangle^r+ \langle
\hat{b}^{\dagger}\hat{b}\rangle^r\right] \nonumber\\
(\omega-\omega_0+i\eta)\langle \hat{b}\hat{b}\rangle^r & =
\Sigma^r(\omega)\left[\langle \hat{b}\hat{b}\rangle^r+ \langle
\hat{b}^{\dagger}\hat{b}\rangle^r\right],
\end{align}
%\begin{align}\label{eq}
%\langle
%&bb^{\dagger}\rangle^r(\omega)=\frac{1}{\omega-\omega_0-\Sigma^r(\omega)+i\eta}\nonumber\\
%\langle
%&b^{\dagger}b\rangle^r(\omega)=-\frac{1}{\omega+\omega_0+\Sigma^r(\omega)+i\eta}\nonumber\\
%&\langle
%bb\rangle^r(\omega)=-\frac{\Sigma^r(\omega)}{(\omega-\omega_0+i\eta)
%(\omega+\omega_0+\Sigma^r(\omega)+i\eta)}\nonumber\\
%&\langle
%b^{\dagger}b^{\dagger}\rangle^r(\omega)=-\frac{\Sigma^r(\omega)}{(\omega+\omega_0+i\eta)
%(\omega-\omega_0-\Sigma^r(\omega)+i\eta)},
%\end{align}
where the retarded self-energy is given by
\begin{equation}\label{self}
\Sigma^r(\omega)=\frac{(MI_0)^2}{\hbar^2}\sum_{j\in
L,R}g_j^2\left(\frac{1}{\omega-\omega_j+i\eta}-\frac{1}{\omega+\omega_j+i\eta}\right).
\end{equation}
The reservoir admittances are given by the Kubo formula
\begin{align}\label{kubo}
Y_L(\omega)&=\frac{i\langle
\hat{i}_L\hat{i}_L\rangle^r(\omega)}{\hbar\omega}\nonumber\\
&=\frac{i}{\hbar\omega}\sum_{j\in
L}g_j^2\left(\frac{1}{\omega-\omega_j+i\eta}-\frac{1}{\omega+\omega_j+i\eta}\right),
\end{align}
so the self-energy is related to the reservoir admittances through
relation
\begin{equation}\label{imp}
\Sigma^r(\omega)=-\frac{iM^2I_0^2\omega}{\hbar}\left[Y_L(\omega)+Y_R(\omega)
\right].
\end{equation}
The algebraic system (\ref{eqof3}) is readily solved yielding
%\begin{align}\label{sol}
%&\langle
%bb^{\dagger}\rangle^r=\frac{-\omega+\omega_0-i\eta+\Sigma^r}
%{(\omega-\omega_0+i\eta)(\omega+\omega_0+i\eta)-2\omega_0\Sigma^r}\nonumber\\
%&\langle b^{\dagger}b\rangle^r=\frac{\omega+\omega_0+i\eta+\Sigma^r}
%{(\omega-\omega_0+i\eta)(\omega+\omega_0+i\eta)-2\omega_0\Sigma^r}\nonumber\\
%&\langle b^{\dagger}b^{\dagger}\rangle^r=\frac{-\Sigma^r}
%{(\omega-\omega_0+i\eta)(\omega+\omega_0+i\eta)-2\omega_0\Sigma^r}\nonumber\\
%&\langle bb\rangle^r=\frac{-\Sigma^r}
%{(\omega-\omega_0+i\eta)(\omega+\omega_0+i\eta)-2\omega_0\Sigma^r}.
%\end{align}
%\begin{align}\label{retrat}
%\langle\hat{I}\hat{I}\rangle^{r}(\omega)&=I_0^2\frac{1}{\frac{(\omega-\omega_0+i\eta)(\omega+\omega_0+i\eta)}
%{2\omega_0}-\Sigma^r}=\nonumber\\
%&\frac{-i\hbar\omega}{F(\omega+i\eta)+M^2\omega^2\left(Y_L(\omega)+Y_R(\omega)\right)},
%\end{align}
\begin{equation}\label{retrat}
\langle\hat{I}\hat{I}\rangle^{r}(\omega)=
\frac{-i\hbar\omega}{\omega
F(\omega+i\eta)+M^2\omega^2\left(Y_L(\omega)+Y_R(\omega)\right)},
\end{equation}
where we have introduced shorthand $F(\omega)\equiv (i\hbar/I_0^2)
(\omega^2-\omega_0^2)/(2\omega_0)$. By extracting the imaginary part
of Eq.~(\ref{retrat}) and recalling the relation
$-2\mathrm{Im}\langle\hat{I}\hat{I}\rangle^{r}(\omega)=S_I(\omega)-S_I(-\omega)$
%\begin{align}\label{imret}
%&S_I(\omega)-S_I(-\omega)=-2\mathrm{Im}\langle\hat{I}\hat{I}\rangle^{r}(\omega)=\nonumber\\
%&\frac{2\hbar\omega\left(\mathrm{Re}[Y_L(\omega)]+\mathrm{Re}[Y_R(\omega)]\right)M^2\omega^2}
%{\left(\mathrm{Im}[F(\omega)]-M^2\omega^2\left(\mathrm{Im}[Y_L(\omega)]+\mathrm{Im}[Y_R(\omega)]\right)\right)^2+
%\left(M^2\omega^2\left(\mathrm{Re}[Y_L(\omega)]+\mathrm{Re}[Y_R(\omega)]\right)\right)^2}.
%\end{align}
we find
%\begin{align}\label{result}
%J=\int_0^{\infty}\frac{d\omega}{2\pi}&\frac{4\hbar\omega^3\mathrm{Re}[Z_L(\omega)^{-1}]\mathrm{Re}[Z_R(\omega)^{-1}]M^4\omega^2}
%{\left(\mathrm{Im}[F(\omega)]-M^2\omega^2\left(\mathrm{Im}[Z_L(\omega)^{-1}]+\mathrm{Im}[Z_R(\omega)^{-1}]\right)\right)^2+
%\left(M^2\omega^2\left(\mathrm{Re}[Z_L(\omega)^{-1}]+\mathrm{Re}[Z_R(\omega)^{-1}]\right)\right)^2}\times\nonumber\\
%&\times\left(n_L(\omega)-n_R(\omega)\right),
%\end{align}
%or the same thing more compactly
\begin{align}\label{resultshort}
J=\int_0^{\infty}\frac{d\omega}{2\pi}&\frac{4\hbar\omega^5M^4\mathrm{Re}[Y_L(\omega)]\mathrm{Re}[Y_R(\omega)]}
{\left|\omega F(\omega)-M^2\omega^2(Y_L(\omega)+Y_R(\omega))\right|^2}\times\nonumber\\
&\times\left[n_L(\omega)-n_R(\omega)\right].
\end{align}
The result (\ref{resultshort}) is valid for arbitrary reservoirs
admittances, as long as the intermediate circuit can be described as
an oscillator. It can be shown that
%for all values of physical
%parameters
the upper limit of heat current (\ref{resultshort}) is given by the
universal single-channel heat conductance quantum
$G_Q=\pi^2k_B^2T/3h$,
\cite{pendry,meschke,schmidt,ojanen,schwab,rego}, an exact parallel
with the electrical single-channel conductance $G_0=2e^2/h$, which
follows from the Meir-Wingreen formula for a noninteracting resonant
level model at resonance \cite{haug}.

\begin{figure}[ht]
\centering
\includegraphics[width=0.90\columnwidth]{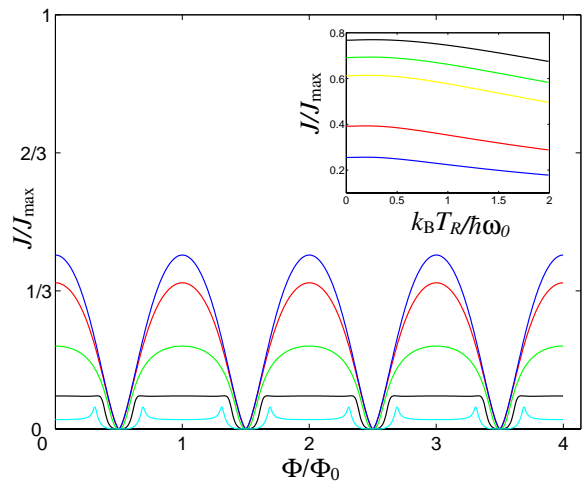}
\caption{Heat flow through the SQUID structure as a function of
applied flux $\Phi$, corresponding to parameters $M^2\omega_0/LR=1$,
$\omega_0/\omega_R=1$ and $T_R=T_L/2=\hbar\omega_0$. The different
curves correspond to different reservoir $Q$ values, $Q=0$ (blue),
$Q=0.1$ (red), $Q=0.5$ (green), $Q=2$ (black) and $Q=10$ (cyan). The
heat flux is normalized with respect to the universal single-channel
maximum value
$J_{\mathrm{max}}=\frac{\pi^2k_{\mathrm{B}}^2}{6h}(T_L^2-T_R^2)$. In
the inset the heat flux is plotted as a function of the temperature
of the right reservoir corresponding to the parameters
$\omega_0/\omega_R=1$, $T_L=2\hbar\omega_0$, $Q=0.1$ and $\Phi=0$
mod $\Phi_0$. The different curves correspond to
$M^2\omega_0/LR=0.5$ (blue), $M^2\omega_0/LR=1$ (red),
$M^2\omega_0/LR=3$ (yellow), $M^2\omega_0/LR=5$ (green) and
$M^2\omega_0/LR=10$ (black).} \label{heatflux}
\end{figure}

In the following we assume that the reservoirs are identical and
consist of a resistor, a capacitor and an inductor in series. Thus
the admittances are can be written as
$Y_L(\omega)=Y_R(\omega)=\frac{R^{-1}}{1+iQ(\omega/\omega_R-\omega_R/\omega)}$,
where $R$, $Q$ and $\omega_R$ are an effective resistance, a quality
factor and a resonance frequency of the reservoir.  The net heat
flow is determined by five dimensionless parameters
$M^2\omega_0/LR$, $\omega_0/\omega_R$, $Q$,
$\hbar\omega_0/k_{\mathrm{B}}T_L$ and
$\hbar\omega_0/k_{\mathrm{B}}T_R$ where $T_{L}$ and $T_R$ are
temperatures of the reservoirs. By externally controlling the
parameters of the middle circuit it is possible to tune the heat
flow through the structure. A practical way to realize this is to
employ a dc-SQUID as a middle circuit and apply a magnetic flux
$\Phi$ through it (see Fig.~\ref{res} b)).
At low temperatures the system can be modeled by an LC-oscillator
with a tunable inductance $L(\Phi)\approx
L/|\mathrm{cos}(\pi\Phi/\Phi_0)|$ where $\Phi_0=h/2e$ is the flux
quantum and $L$ is determined by the Ambegaokar-Baratoff critical
current \cite{tinkham}. The LC-oscillator description of the SQUID
circuit is expected to be accurate within realistic parameter values
at the experimentally verified crossover temperature
$T_{\mathrm{cr}}\sim 100$ mK below which the photonic thermal
conductance should dominate \cite{meschke}. A numerical evaluation
of Eq.~(\ref{resultshort}) is presented in Fig.~\ref{heatflux},
which shows the heat flow as a function of the external bias flux
$\Phi$. The maximum flow is obtained at at integer values of
$\Phi/\Phi_0$, whereas at half-integer values the reservoirs are
thermally decoupled. At low quality factors the reservoirs are more
efficiently matched and the system has a better thermal coupling. By
increasing the coupling parameter $M^2\omega_0/LR$ the maximum value
of the heat flow could be enhanced closer to the single-channel
maximum value.

In summary, we have studied photon heat transport in nanoelectronic
circuits based on a novel Meir-Wingreen-Landauer formula in a
two-terminal geometry. This formula expresses the heat current in
terms of the admittances of the heat reservoirs, and the noise power
of the intermediated mesoscopic circuit.  The formula can serve as a
starting point for an analysis of photon heat transport in a wide
range of applications.  As an example, we present an exact solution
to the transport problem in the case of an electromagnetic resonator
playing the role of the mediating circuit. We propose a new device
concept, a mesoscopic photon heat transistor, where the heat current
through the structure can be strongly modulated by the external
magnetic flux through a dc-SQUID loop.

We would like thank Antti Niskanen, Jukka Pekola and Tero Heikkil\"a
for valuable discussions.

\begin{thebibliography}{99}
\bibitem{note}A direct coupling between the
reservoirs could easily be included in the formalism, but since it
is can be made small by the device geometry, we will not discuss it
here.
\end{thebibliography}
\end{document}